\documentclass[reprint,aps,prb,twocolumn,floats,footinbib,superscriptaddress]{revtex4-1}

\usepackage{bm} 
\usepackage{graphicx}
\usepackage{amsmath}
\usepackage{mathtools}
\usepackage{amsfonts}
\usepackage{amssymb}
\usepackage{bbold}
\usepackage{times} 
\usepackage[T1]{fontenc}
\usepackage{upgreek}
\usepackage{outlines}
\usepackage{latexsym}
\usepackage{textcomp} 
\usepackage{verbatim}
\usepackage{fancyvrb} 
\usepackage[ddmmyyyy,hhmmss]{datetime}
\usepackage[usenames]{color}
\definecolor{R}{RGB}{252,0,0}
\usepackage{hyperref}
\usepackage[utf8]{inputenc}
\newcommand{\angstrom}{\text{\normalfont\AA}}
\usepackage{siunitx}

\newcommand{\prom}{\text{\textperthousand}}

\newcommand{\veps}{\varepsilon}

\newcommand{\vareps}{\varepsilon}
\newcommand{\beq}{ \begin{equation} }
\newcommand{\eeq}{\end{equation}}

\newcommand{\closure}[2][3]{%
{}\mkern#1mu\overline{\mkern-#1mu#2}} 
\newcommand{\bvalue}{${b=(-0.94\pm0.11)}$~eV}
\newcommand{\Gvalue}{${G_{11}=(72.2\pm1.9)}$~neV}
\newcommand{\Cratio}{$2C_{12}/C_{11}=1.39886$}
\newcommand{\SORTNOOPCYR}[1]{}
\newcommand\Tstrut{\rule{0pt}{2.6ex}}         

\def \PAST{Institute of Experimental Physics, Faculty of Physics,
University of Warsaw, Pasteura 5, 02-093 Warsaw, Poland}
\def \IFPAN{Institute of Physics, Polish Academy of Sciences, Aleja
Lotnik\'ow 32/36, 02-668 Warsaw, Poland}

\begin{document} 

\title{Angle-resolved optically detected magnetic resonance as a tool for strain
determination in nanostructures} 

\author{A.
\surname{Bogucki}}\email{Aleksander.Bogucki@fuw.edu.pl}\affiliation{\PAST}
\author{M.
\surname{Goryca}}\affiliation{\PAST}
\author{A.
\surname{\L{}opion}}\affiliation{\PAST}
\author{W.
\surname{Pacuski}}\affiliation{\PAST}
\author{K.
\surname{Po{\l}czy\'nska}}\affiliation{\PAST}
\author{J.
\surname{Domaga\l{}a}}\affiliation{\IFPAN}
\author{M.
\surname{Tokarczyk}}\affiliation{\PAST}
\author{T.
\surname{F\k{a}s}}\affiliation{\PAST}
\author{A.
\surname{Golnik}}\affiliation{\PAST}
\author{P.
\surname{Kossacki}}\affiliation{\PAST}


\date{\today}

\begin{abstract} 
In this paper, we apply the angle-resolved Optically Detected Magnetic Resonance (ODMR) technique to study series of strained (Cd, Mn)Te/(Cd, Mg)Te quantum wells (QWs) produced by molecular beam epitaxy. By analyzing characteristic features of ODMR angular scans, we determine strain-induced axial-symmetry spin Hamiltonian parameter $D$ with neV precision. Furthermore, we use low-temperature optical reflectivity measurements and X-ray diffraction scans to evaluate the local strain present in QW material. In our analysis, we take into account different thermal expansion coefficients of GaAs substrate and CdTe buffer. The additional deformation due to the thermal expansion effects has the same magnitude as deformation origination from the different compositions of the samples.  Based on the evaluated deformations and values of strain-induced axial-symmetry spin Hamiltonian parameter $D$, we find strain spin-lattice coefficient \Gvalue ~for Mn$^{2+}$ in CdTe and shear deformation potential \bvalue ~for CdTe.
\end{abstract}


\maketitle

\section{Introduction}
One of the critical factors influencing the electronic and optoelectronic devices' performance is local strain distribution. For example, the spin relaxation time -- a crucial parameter for potential spintronic devices -- strongly depends on spin-lattice coupling and local strain distribution. The latter is particularly non-trivial in complex nanostructures composed of many different materials, e.g., in quantum wells (QW). Such structures can be produced with several different growth techniques like metalorganic vapor-phase epitaxy (MOVPE) or molecular beam epitaxy (MBE). Historically, the development of semiconductors' growth methods was accompanied by independent improvement of materials characterization techniques, which resulted in a comprehensive knowledge foundation for future research and advanced applications. Naturally, new characterization techniques combined with new growth methods offer fresh insights into semiconductor physics and often shed new light on some previously determined material parameters. For example, this was the case of GaAs crystals widely used as substrates for complex semiconductor structures. In 1959 Kolm et al. \cite{Kolm_1957_PR} used X-ray diffraction to determine the lattice constant of the GaAs crystals produced with the Bridgman–Stockbarger method. However, crystals grown nowadays with the use of other techniques like Horizontal Bridgman (HB), Liquid Encapsulated Czochralski (LEC), Vertical Gradient Freeze (VGF), or epitaxial methods resulting in different concentrations of impurities systematically exhibit different lattice constant \cite{Bassignana_1997_JoCG, Usuda_1996_JoAP}.

Finding a particular material parameter for materials obtained by a specific growth method is critical for real multilayer devices. Determining some structural parameters is particularly challenging in small-volume structures (e.g. single QWs). A small amount of material often excludes many characterization techniques like X-ray diffraction which is a standard method of measuring the strain in semiconductor nanostructures. A possible solution to this issue is an incorporation of a small amount (below 1\%) of magnetic ions into the structure and determination of local strain exploiting the coupling between those ions and the strain of the crystal lattice. In many cases (like multiple QWs structures), determining the spin Hamiltonian parameters describing such coupling may be done with electron paramagnetic resonance (EPR) techniques \cite{Qazzaz_1995_SSC}. However, in thin structures, the number of spins is insufficient to produce a detectable EPR signal. This problem can be overcome by using the optically detected magnetic resonance (ODMR) technique which exploits strong exchange interaction between magnetic ions and photo-generated carriers and facilitates detection of the paramagnetic resonance of significantly smaller amount of magnetic ions.

\begin{figure}[h]
\begin{center}
\includegraphics[width=85.9mm]{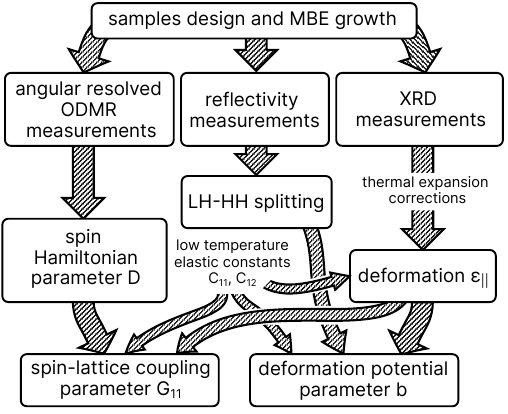}
\end{center}
\caption[]{The schematic overview of this work. There are three main branches related to the experimental techniques used in this paper: angular resolved ODMR measurements, reflectivity measurements, and XRD measurements. By combining the results of these methods we determine the spin-lattice coupling parameter $G_{11}$ and deformation potential parameter $b$ for Mn$^{2+}$ ion in CdTe.}
\label{fig:Paper_scheme}
\end{figure}


In this paper, we present a comprehensive analysis of the local strain present in series of (Cd, Mn)Te/(Cd, Mg)Te QWs. We combine data obtained from angle-resolved ODMR, low-temperature reflectance measurements, and X-ray diffraction scans. The scheme presented in figure \ref{fig:Paper_scheme} shows how we address problems formulated in the previous paragraph. We design a series of (Cd, Mn)Te/(Cd, Mg)Te quantum wells samples fabricated by MBE method that have different deformation in the QW layer by design. The QW material is doped with a small amount of manganese ($<1$\%) which exhibits paramagnetic resonance. Using angle-resolved ODMR, we determine strain-induced axial-symmetry spin Hamiltonian parameter ($D$) with high accuracy. Independently, we perform XRD measurements on the least and the most strained samples -- as only for these samples the XRD method yields reliable deformation value. After considering additional deformation that originates from different thermal expansion coefficients of the substrate and QW material, we deduce deformation values for all samples. To check if other samples also follow the design scheme, we perform low-temperature reflectivity measurements. We determine energy splitting between light-hole (LH) exciton transition and heavy-hole (HH) exciton transition from reflectivity spectra.  We find that the HH-LH splitting for the whole series of samples exhibits monotonic behavior - thus validating the designed distribution of deformation.  As the ODMR measurements are taken at cryogenic temperatures, we evaluate CdTe elastic stiffness constants $C_{11}$ and $C_{12}$ at 1.6~K from available literature data. Then from obtained deformation values and corresponding strain-induced axial-symmetry spin Hamiltonian parameter $D$ combined with elastic constants, we calculate the spin-lattice coupling coefficient $G_{11}$ for manganese in cadmium telluride.  The spin-lattice coupling coefficient $G_{11}$ value obtained in this work diverges from previously reported in ref. \onlinecite{Causa_1980_PL}. Furthermore, we use HH-LH splittings and deformation data to find deformation potential $b$. The obtained value of $b$ is similar to ref's \onlinecite{MerledAubigne_1990_JCG} one but different from values reported in refs. \onlinecite{Thomas_1961_JoAP,Mathieu_1988_PRB,Peyla_1992_PRB}. 

Our findings could be especially useful in all studies involving spin-related phenomena in CdTe-based MBE-grown nanostructures and devices including multiple-QWs and quantum dots \cite{Cherbunin_2020_PRB,Lafuente-Sampietro_2017_PRBa,Dinu_2017_PSSB,Moldoveanu_2016_PRB,Lafuente-Sampietro_2016_PSSC,Lafuente-Sampietro_2015_PRBa,Goryca_2015_PRB,Cronenberger_2015_NC,Kobak_2014_NC,Varghese_2014_PRB,Besombes_2014_PRB,Goryca_2014_PRLa,Cronenberger_2013_PRL,Jamet_2013_PRB,Chen_2013_PRB,Besombes_2012_PRB,Goryca_2009_PRL,Goryca_2009_PRLa,LeGall_2009_PRLa}.

\section{Experiment and results}

\subsection{Samples}
\begin{figure}[h]
\begin{center}
\includegraphics[width=80mm]{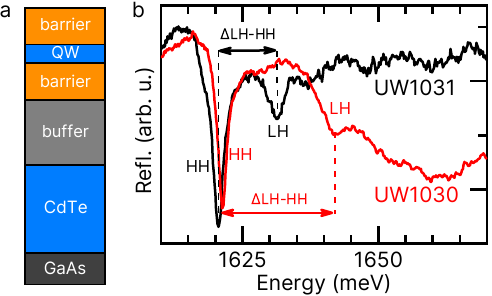}
\end{center}
\caption[]{(Color online) (a) An schematic of a representative sample structure used in the
whole sample series used in this article. The buffer layer was made of
(Cd, Mg)Te where magnesium content was varying from 0\% up to approx. 30\%. 
Barriers were made of (Cd, Mg)Te while quantum well (QW) mas made of
(Cd, Mn)Te with manganese concentration below 0.5\%. (b) Reflectivity
spectra of two example samples. Sample UW1030 has a higher concentration of
magnesium in the buffer layer than UW1031 and thus exhibits larger strain
which results in larger energy splitting between heavy-hole exciton (HH)
and light-hole exciton (LH).}
\label{fig:Samples_Reflect}
\end{figure}

The samples containing QWs used in this work were produced by the Molecular Beam
Epitaxy (MBE) technique. The scheme of samples is presented in figure
\ref{fig:Samples_Reflect}a.  In order to obtain structures with a different
strain, we have changed the content of magnesium in the buffer layer and the barrier layer (see table \ref{tab:samples}). The magnesium content was deduced based on magnesium fluxes and calibration obtained by the reflectance measurements \cite{Hartmann_1996_JAP, LeBlanc_2017_JoEM}.  The nominal width of (Cd, Mn)Te QW was
10~nm. The value of the QW width was chosen to ensure sufficient confinement together with narrow excitonic features in the reflectivity spectra, as well as it is well below the critical thickness of the CdTe lattice relaxation \cite{Cibert_1990_APL,Cibert_1991_SaM}.  The manganese content was determined by magnetooptical measurements of the giant Zeeman splitting and fitting the modified Brillouin function\cite{Gaj_1994_PRB}. The thickness of the (Cd, Mg)Te barriers was 50~nm, the (Cd, Mg)Te buffer layer was 2000~nm (far above the critical thicknesses of the (Cd, Mg)Te\cite{Waag_1993_JoCGb}) and the additional CdTe layer below the buffer was 4000~nm thick. This additional layer ensured the isolation from the GaAs substrate.  To confirm that we obtained designed strain in produced samples we have performed reflectivity measurements (see figure \ref{fig:Samples_Reflect}b). The identification of the spectrum features was based on standard magneto-optical measurements (not shown)\cite{Kossacki_1997_SSC,Kossacki_2003_JPCM,Kossacki_2004_PRB}. The energy splitting between light-hole exciton and heavy-hole exciton visible in spectroscopic measurements is a linear function of the strain present in the QW layer \cite{MerledAubigne_1990_JCG,Allegre_1990_JoCG,Boley_1993_JoAP,Bhattacharjee_1998_JoCG}. Therefore, as the samples are similarly designed, the light-hole heavy-hole splitting gives a hint about the existing strain.

\begin{table}[h]
\caption{Composition of samples used in this article, deformation present in QW calculated for 300~K without correction for different thermal expansion coefficients of CdTe and GaAs, deformation present in QW calculated for 1.6~K present in QW layer with thermal-expansion-coefficients-difference correction.}
\begin{tabular}{cS[table-format=2.1]S[table-format=2.1]S[table-format=1.2]S[table-format=2.3]S[table-format=2.3]}
\toprule
           &  {buffer}  & {barrier}  & {QW}  & ${\veps^{T=300~K}_{\parallel \mathrm{~w/o~corr.}}}$
& ${\veps^{T=1.6~K}_{\parallel \mathrm{~corr.}}}$   \Tstrut \\
{sample no.} &  {Mg (\%)} &  {Mg (\%)} &  {Mn (\%)} & {(\prom)} & {(\prom)}    \\
\hline
UW1029 & 30.7 & 30.7 & 0.30 & -2.870 & -3.512  \\
UW1030 & 21.2 & 21.2 & 0.31 & -1.959 & -2.601  \\
UW1028 & 16.4 & 16.4 & 0.14 & -1.538 & -2.180  \\
UW0677 & 15.2 & 15.2 & 0.26 & -1.396 & -2.038  \\
UW1050 & 8.5  & 16.4 & 0.15 & -0.780 & -1.422  \\
UW1031 & 0    & 21.2 & 0.26 & 0.058  & -0.584  \\
UW0676 & 0    & 17.3 & 0.27 & 0.060  & -0.582 \\\hline \hline
\end{tabular}
\label{tab:samples}
\end{table}

\begin{figure}[h]
\begin{center}
\includegraphics[width=83mm]{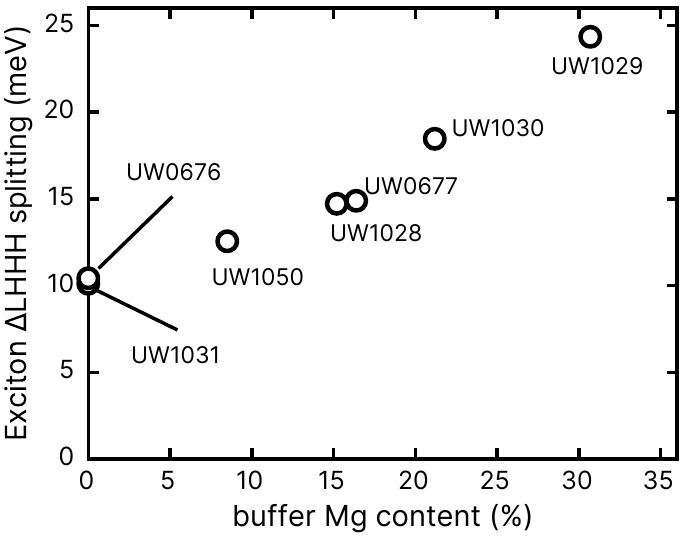}
\end{center}
\caption[]{The energy difference between the lowest light-hole excitonic state and lowest heavy-hole excitonic state ($\Delta$LHHH splitting) plotted versus magnesium content in the buffer layer for all samples from table \ref{tab:samples} confirms that the growth process was fully controlled.}
\label{fig:HHLH_vs_Mg}
\end{figure}

Figure \ref{fig:HHLH_vs_Mg} shows heavy-hole exciton light-hole exciton splitting obtained as a function of magnesium content in the buffer layer for all samples. The monotonic behavior of extracted data confirms that samples follow the design -- the higher magnesium content, the larger light-hole heavy-hole splitting.

\subsection{XRD measurements}
\label{label:XRD}
\begin{figure}
\begin{center}
\includegraphics[width=85.9mm]{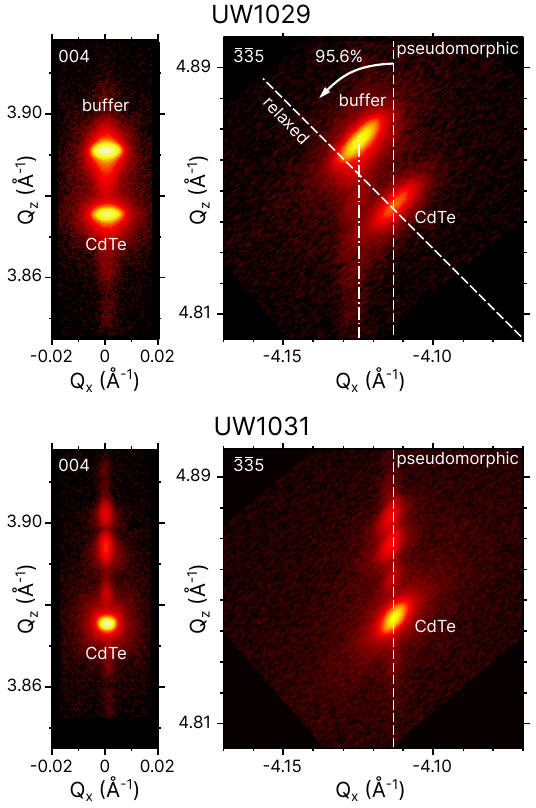}
\end{center}
\caption[]{(Color online) Upper panel shows XRD maps with 004 (upper left) and $\closure[1]{3}\closure[1]{3}5$ (upper right) reflections obtained for 
high-strain sample (UW1029). There are two strong peaks attributed to CdTe and to (Cd, Mg)Te buffer layer. Analysis of  $\closure[1]{3}\closure[1]{3}5$ map confirms that strain existing in the QW is induced by buffer layer, as weak satellite peaks are present along vertical line meaning pseudomorphic growth with buffer lattice constant. The lower panel presents maps from the same region of reciprocal space (lower left panel the 004 reflection, lower right panel for $\closure[1]{3}\closure[1]{3}5$ reflection) for the low-strain sample (UW1031). All peaks present in $\closure[1]{3}\closure[1]{3}5$-map appear along the horizontal line above CdTe peak, confirming the pseudomorphic character of growth with CdTe lattice constant.}
\label{fig:XRD_maps}
\end{figure}

In order to independently calibrate deformation, we have performed room-temperature X-ray diffraction (XRD) measurements of the two extreme samples: UW1029 (highest strain) and UW1031 (lowest strain) -- see figure \ref{fig:XRD_maps}. By analyzing the 004 and the $\closure[1]{3}\closure[1]{3}5$ reflections, we extract lattice constants of CdTe separation layer (about a$=6.481$~\angstrom) and the strained layers. Throughout this work, for the calculations we use the following lattice constants of pure materials at room temperature: a$_{\mathrm{CdTe}}=6.481$~\angstrom, a$_{\mathrm{MnTe}}=6.337$~\angstrom, a$_{\mathrm{MgTe}}=6.419$~\angstrom~and a$_{\mathrm{GaAs}}=5.6535$~\angstrom.~Obtained results agree with the sample design.

In XRD maps of the high-strain sample (UW1029), there are two strong
reflections -- one corresponding to 4~$\upmu$m CdTe layer, and the second
one to 2~$\upmu$m (Cd, Mg)Te buffer layer. Positions of the reflections give the lattice
constants of CdTe and buffer at room temperature (where $_{\parallel}$ and $_{\perp}$ denote directions parallel and perpendicular to the samples surface, respectively):
$a^{\textrm{CdTe}}_{\parallel}=(6.4807\pm 0.0006)$~\angstrom, ~
$a^{\textrm{CdTe}}_{\perp}=(6.4862\pm 0.0001)$~\angstrom, ~
$a^{\textrm{buff}}_{\parallel}=(6.4616\pm 0.0003)$~\angstrom, ~
$a^{\textrm{buff}}_{\perp}=(6.4595\pm 0.0001)$~\angstrom.~From this values and using a
formula $$a_{\textrm{relax}} = (C_{11}a_{\perp}+2C_{12}a_{\parallel})/(
C_{11}+2C_{12}) $$ we can calculate lattice constant of relaxed material and 
finally obtain deformation of the buffer layer
$\varepsilon_{\parallel}=-0.3565\prom$. The fact that nominally relaxed material has nonzero deformation is explained in following sections. The $C_{12}$ and $C_{11}$ are elastic stiffness constants of CdTe. More details about evaluation of elastic stifness constants at different temperatures are provided in the Appendix. The analysis of $\closure[1]{3}\closure[1]{3}5$ reflection shows that the buffer layer is not fully relaxed. The relaxation degree of the buffer layer in UW1029 sample is $95.6\%$. 

The situation is easier to interpret in the case of the low-strain sample (UW1031) where on top of the separation layer made of CdTe there is a buffer layer also made from CdTe. In principle, the lattice constant of the thin, strained barrier (thickness well below critical thickness) should be matched to the lattice constant of the thick CdTe buffer. Therefore lattice constant of the QW should be close to the lattice constant of CdTe. The obtained values of lattice constants for the UW1031 sample from XRD measurements are:
$a^{\textrm{CdTe+buffer}}_{\parallel}=(6.4807\pm 0.0006)$~\angstrom, ~
$a^{\textrm{CdTe+buffer}}_{\perp}=(6.4867\pm 0.0001)$~\angstrom, ~
$a^{\textrm{2nd}}_{\parallel}=(6.4797\pm 0.0008)$~\angstrom, ~
$a^{\textrm{2nd}}_{\perp}=(6.4549\pm 0.0001)$~\angstrom. From this, we can calculate $\varepsilon_{\parallel}=-0.389\prom$ which in the case of the UW1031 sample is also the deformation present in QW at room temperature. 

The influence of temperature is critical in the correct determination of deformation in heterogenic structures \cite{Horning_1987_APL, Tatsuoka_1989_JoAP, Tatsuoka_1990_JoAP, Tatsuoka_1991_TSF}  which is the case of our sample, as the used materials exhibit different temperature expansion coefficients. There are three distinctive temperature points that should be considered in the context of this work. i) The growth temperature at which CdTe layers were deposited in MBE.  We have estimated this temperature to be 552~K -- see Appendix. ii) The room temperature ($300$~K) at which XRD measurements were taken. During the cooling after the growth process, an additional compressive deformation builds up in the QW layer, as the GaAs substrate shrinks more than CdTe during the cooling. So at room temperature there is an additional deformation originating from the different shrinking of the QW layer and the substrate layer that is visible in XRD measurements: $\varepsilon^{\textrm{552K}\rightarrow\textrm{300K}}_{\parallel}=-0.389\prom$.  iii) The pumped liquid helium temperature ($\approx1.6$~K) at which reflectivity and ODMR measurements were performed. The GaAs temperature expansion coefficient differs from that of the CdTe (see figure \ref{fig:app_thermal} in Appendix). Moreover, the magnitude of that difference changes as the temperature decreases finally resulting in an additional compression which gives contribution to the deformation $\varepsilon^{\textrm{300K}\rightarrow\textrm{1.6K}}_{\parallel}=-0.253\prom$. Taking into account all of those contributions up together with starting value $\veps^{T=300~K}_{\parallel \mathrm{~w/o~corr.}}=0.058\prom$ we finally obtain deformation in the CdTe QW layer at 1.6~K for the UW1031 sample $\varepsilon_{\parallel}=-0.584\prom$.

\begin{figure}
\begin{center}
\includegraphics[width=85.9mm]{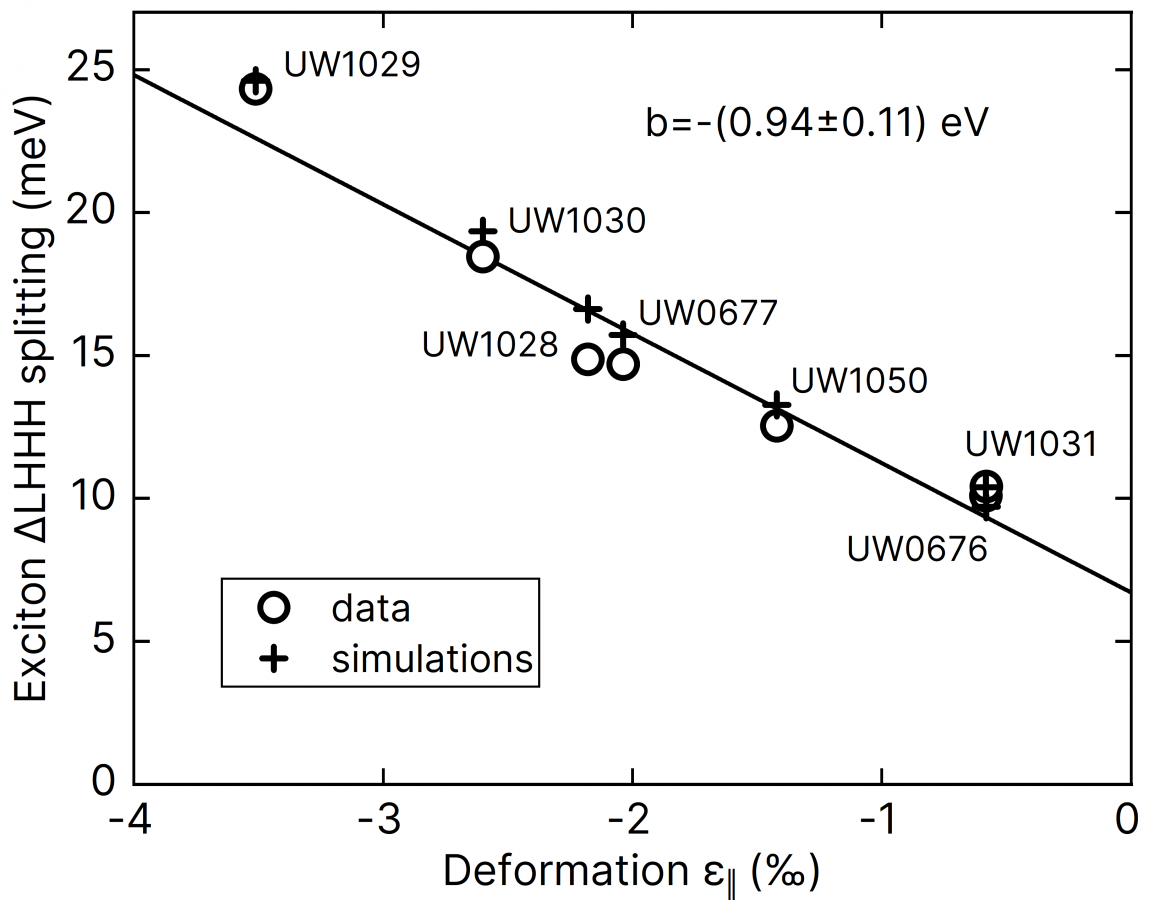}
\end{center}
\caption[]{The energy difference between the lowest light-hole excitonic state and lowest heavy-hole excitonic state ($\Delta$LHHH splitting) plotted as a function of deformation
$\vareps_{\parallel}$ at $1.6$~K for analyzed quantum well samples. The empty circles represent measured data, the black crosses are results of numerical simulations described in Appendix \ref{label:XRD}. The black line is a result of a linear fit to the data points. The shear deformation potential  \bvalue ~is calculated from the slope of an obtained line (equation \ref{eq:Delta_LHHH}). }   
\label{fig:HH_LH_eps}
\end{figure}

Now, using above calibration and taking into account an additional temperature-originating deformation it is possible to present light-hole-heavy-hole splitting as a function of deformation (figure \ref{fig:HH_LH_eps}). We obtain band shear deformation potential \bvalue ~using the formula \cite{Chuang_2009_book}:
\begin{equation}
    \Delta \mathrm{LHHH}  = 2b \left (1+\frac{2C_{12}}{C_{11}} \right) \varepsilon_{\parallel}+ \mathrm{const.}
\label{eq:Delta_LHHH}
\end{equation}

The shear deformation potential obtained in this work is close to the value $(-1.05\pm0.01)$~eV obtained by Merle d`Aubign\'e et al.  \cite{MerledAubigne_1990_JCG}. The other experimentally determined value is $b=1.24$~eV by Thomas and coworkers \cite{Thomas_1961_JoAP}. Other values available are given by Peyla et al. \cite{Peyla_1992_PRB}, and by Mathieu et al. $-1.4$~eV \cite{Mathieu_1988_PRB}. However, the two latter ones are based on refs. \cite{Thomas_1961_JoAP, MerledAubigne_1990_JCG}. Namely, the value given by Peyla et al. \cite{Peyla_1992_PRB} is the mean of values given by Thomas and Merle.    The results of calculated light-hole heavy-hole splitting obtained from numerical solving the finite QW model are marked in Figure \ref{fig:HH_LH_eps} with black crosses. More details about QWs simulations are in Appendix.  

\begin{figure}
\begin{center}
\includegraphics[width=85.9mm]{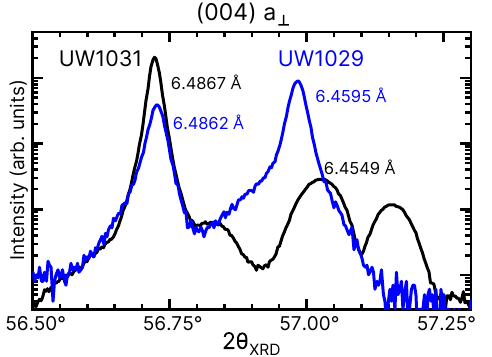}
\end{center}
\caption[]{(Color online) The $2\theta$ XRD scan of the 004 reflection for low-strain (UW1031, black line) and high-strain (UW1029, blue line) samples. Lattice constants calculated from this measurement confirm that obtained strain agrees with the designed one. }
\label{fig:XRD_2theta}
\end{figure}

\subsection{Optically Detected Magnetic Resonance measurements}
\begin{figure}[t]
\begin{center}
\includegraphics[width=0.9\columnwidth]{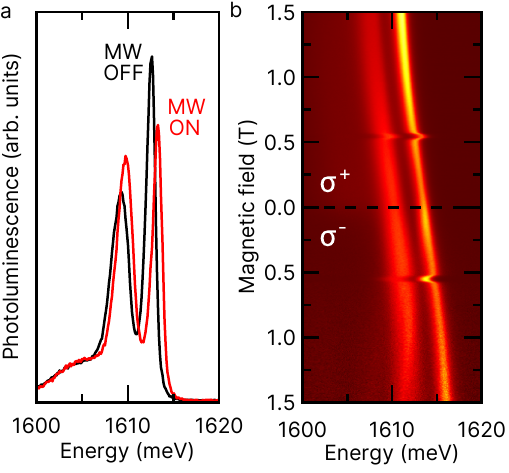}
\end{center}
\caption[]{(Color online) (a) Photoluminescence (PL) spectra of (Cd, Mn)Te/(Cd, Mg)Te quantum well (QW) collected at $1.6$~K at a magnetic field of $0.545$~T with microwave radiation ($15.3$~GHz) turned on (red curve) and off (black curve).  (b) Map of the normalized photoluminescence spectra of the same QW measured for various magnetic fields with present microwave radiation. The energetic position of the PL spectrum follows the modified Brillouin function, except the points at $0.545$~T, where the paramagnetic resonance of Mn$^{2+}$ ions placed in the QW occurs. As a result of the resonance, the giant Zeeman splitting rapidly decreases in those points.}
\label{fig:what_is_ODMR}
\end{figure}

\begin{figure}[hb]
\begin{center}
\includegraphics[width=0.5\columnwidth]{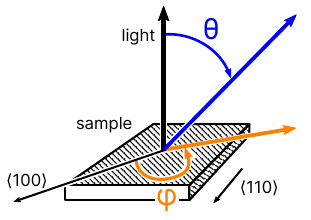}
\end{center}
\caption[]{(Color online) Angles definition of the magnetic field directions. The sample edge is along $\langle$110$\rangle$ direction. The out-of-plane angle $\theta$ is the angle between the samples' growth axis and samples' surface. For example, $\theta=0^{\circ}$ means magnetic field along growth axis in Faraday configuration, whereas $\theta=90^{\circ}$ means in-plane magnetic field.   The in-plane angle $\varphi$ is the angle between $\langle$100$\rangle$ direction and direction of the magnetic
field. The angle $\varphi=45^{\circ}$ means that the in-plane magnetic field was along $\langle$110$\rangle$ direction.}
\label{fig:katy_definicja}
\end{figure}

Optically Detected Magnetic Resonance is a technique that exploits the fact that
the optical properties of the studied material change when paramagnetic
resonance occurs. The microwave absorption at electron paramagnetic resonance (EPR) frequency leads to an increase of Mn-system temperature which results in a decrease of optically-detected magnetization \cite{Gisbergen_1993_PRB,Ivanov_2001_APPA,Godlewski_2002_PSSB,Khachapuridze_2002_APPA,Godlewski_2004_JoAaC,Smith_2005_APL,Godlewski_2008_OM,Baranov_2009_JL,Debus_2016_PRB,O.Tolmachev_2020_N,Shornikova_2020_AN}.  The change in the optical properties of the sample can be detected with various experimental techniques. For example, previous ODMR studies of Diluted Magnetic Semiconductors were based on changes in Faraday rotation \cite{Komarov_1977_JETP}, in the amplitude of
photoluminescence or in the spectral shift of photoluminescence line
\cite{Ivanov_2008_PRB,Ivanov_2010_AMR,
Sadowski_2003_APL, Byszewski_2004_PELSN,Tolmachev_2010_PSSB,Tolmachev_2012_JL,Gurin_2015_JL}.  
In this work, we use the energy position of the neutral exciton emission
line (X) from photoexcited QW to detect local magnetization in the QW layer (fig. \ref{fig:what_is_ODMR}). For optical
excitation, we use $\lambda=647$~nm laser (with a spot diameter of approx. $100\:\upmu$m) in a standard photoluminescence experimental setup. The sample is placed in the optical cryostat equipped with two perpendicular pairs of split superconducting coils. Thus it is possible to obtain up to $3$~T of a magnetic field in any direction in a given plane. In particular, it is possible to continuously change the magnetic field from Faraday to Voigt configuration (scan over the out-of-plane angle $\theta$ in figure \ref{fig:katy_definicja}) or to rotate the magnetic field within the sample plane (scan over the in-plane angle $\varphi$ in figure~\ref{fig:katy_definicja}). 

\begin{figure}[t]
\begin{center}
\includegraphics[width=70.9mm]{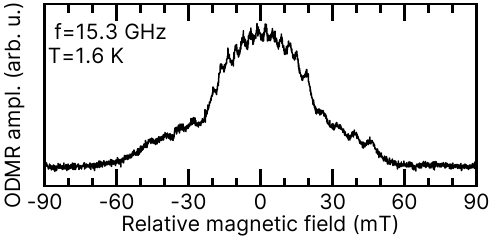}
\end{center}
\caption[]{Optically Detected Magnetic Resonance (ODMR) signal of
(Cd, Mn)Te/(Zn, Mg)Te quantum well obtained as a difference between the energetic position of the quantum well excitonic line measured with microwave radiation and expected position of the line calculated from
fitted modified Brillouin function without microwave radiation.  On the
the horizontal axis, we used the relative magnetic field -- a magnetic field
with a subtracted value of the resonance magnetic field for a given
frequency of microwave radiation. The multiple features visible on the
spectrum are originating from different terms of manganese Hamiltonian
hyperfine structure interaction.}
\label{fig:ODMR_Signal}
\end{figure}

The representative photoluminescence spectrum (PL) for samples used in this
work is presented in figure \ref{fig:what_is_ODMR}a. The PL spectrum
consists of two emission lines --- at lower energies there is a line related to the charged
exciton (CX) and at higher energies, to the neutral exciton (X)\cite{Kossacki_2004_PRB}. During the scan in the magnetic field in Faraday configuration (magnetic field perpendicular to the surface of the sample and parallel
to the optical axis) energetic position of X~line can be described with the
modified Brillouin function \cite{Gaj_1994_PRB} (giant Zeeman effect).
However, in the presence of microwave radiation, a paramagnetic resonance
of manganese ions occurs, at the resonant magnetic field, which results in a
rapid decrease of giant Zeeman splitting (see figure \ref{fig:what_is_ODMR}b with its caption).  Comparison of PL spectra
measured for a resonance magnetic field with and without microwave radiation are presented in figure \ref{fig:what_is_ODMR}a. The difference in the energy position of X~line for these two cases is called further
in the text ''the ODMR amplitude/signal``.  It is important to note that too strong laser excitation can alter the optical properties of QW resulting in a change of the position of the X line. To avoid this effect and ensure that the laser excitation does not influence an amplitude of the Giant Zeeman splitting \cite{Keller_2001_PRB, Seufert_2001_PRL, Yakovlev_2004_PSSC, Hundt_2005_PRB} we use the following procedure to determine an optimal laser excitation power. The magnetic field is set to the value of paramagnetic resonance with microwave radiation turned off. The series of PL spectra are measured for decreasing laser power. The optimal laser power is a power that does not cause an energy shift of the neutral exciton emission line. In this work, the optimal laser power corresponds to the excitation density power $\rho<0.005$~W/cm$^2$, which is even smaller than the low-power regime value presented in ref. \onlinecite{Keller_2001_PRB}.

An example of representative ODMR signal as a function of magnetic field for fixed
microwave frequency is presented in figure~\ref{fig:ODMR_Signal}. The detailed features (multiple lines) visible in the ODMR spectrum originate from interactions of  manganese ion Mn$^{2+}$ and CdTe material in QW and
can be described with spin Hamiltonian \cite{Qazzaz_1995_SSC,
Abragam_2012_EPR}:

\begin{eqnarray}
\nonumber\widehat{H}&=&g_{\mbox{\tiny{Mn}}}\mu_B\mathbf{\hat{B}\hat{S}}+A\mathbf{\hat{I}\hat{S}}+D\left[\hat{S}_z^2-\frac{S(S+1)}{3}\right]+{}\\
\label{eq:hamiltonian}& &{}+E\; R_z(\alpha) \left (  \hat{S}_x^2 -
\hat{S}_y^2  \right ) R_z^{-1}(\alpha)+\\ 
\nonumber&
&{}+\frac{a}{6}\left[\hat{S}_x^4+\hat{S}_y^4+\hat{S}_z^4-\frac{S(S+1)(3S^2+3S-1)}{5}\right].
\end{eqnarray} 

The first term of the Hamiltonian is responsible for Zeeman splitting (where
$g_{\mbox{\tiny{Mn}}}$ is g-factor of manganese ion). The second one is a
hyperfine coupling between the electronic spin and nuclear spin and it
results in splitting the ODMR spectrum into $6$ lines as manganese $2+$ ion has electronic spin $S=5/2$ and nuclear spin $I=5/2$. The third and fourth term comes from electronic quadrupole fine structure present in less than cubic symmetry (strained QW) and can be written in a general form as $\hat{S}\mathbf{\hat{D}}\hat{S}$ where $\mathbf{\hat{D}}$ is a tensor.  However, it can be reduced to only two parameters $D$ and $E$ -- the axial and rhombic zero-field splitting (ZFS) parameters ($D$ is also called the strain-induced axial-symmetry parameter)\cite{Abragam_2012_EPR}. The  $R_z(\alpha)$ allows for the
rotation of the reference frame around the growth axis of the QW.  The last
term describes the crystal-field splitting where $a$ is the zero-field fine
structure splitting parameter for the unstrained CdTe. As a consequence of
the above Hamiltonian, the microwave absorption spectrum of Mn$^{2+}$ ion
consists of $30$ transitions ($5$ electronic spin transitions, each split into
the sextet due to hyperfine interaction).
\subsection{Angular resolved ODMR}

\begin{figure}[!h]
\begin{center}
\includegraphics[width=85.9mm]{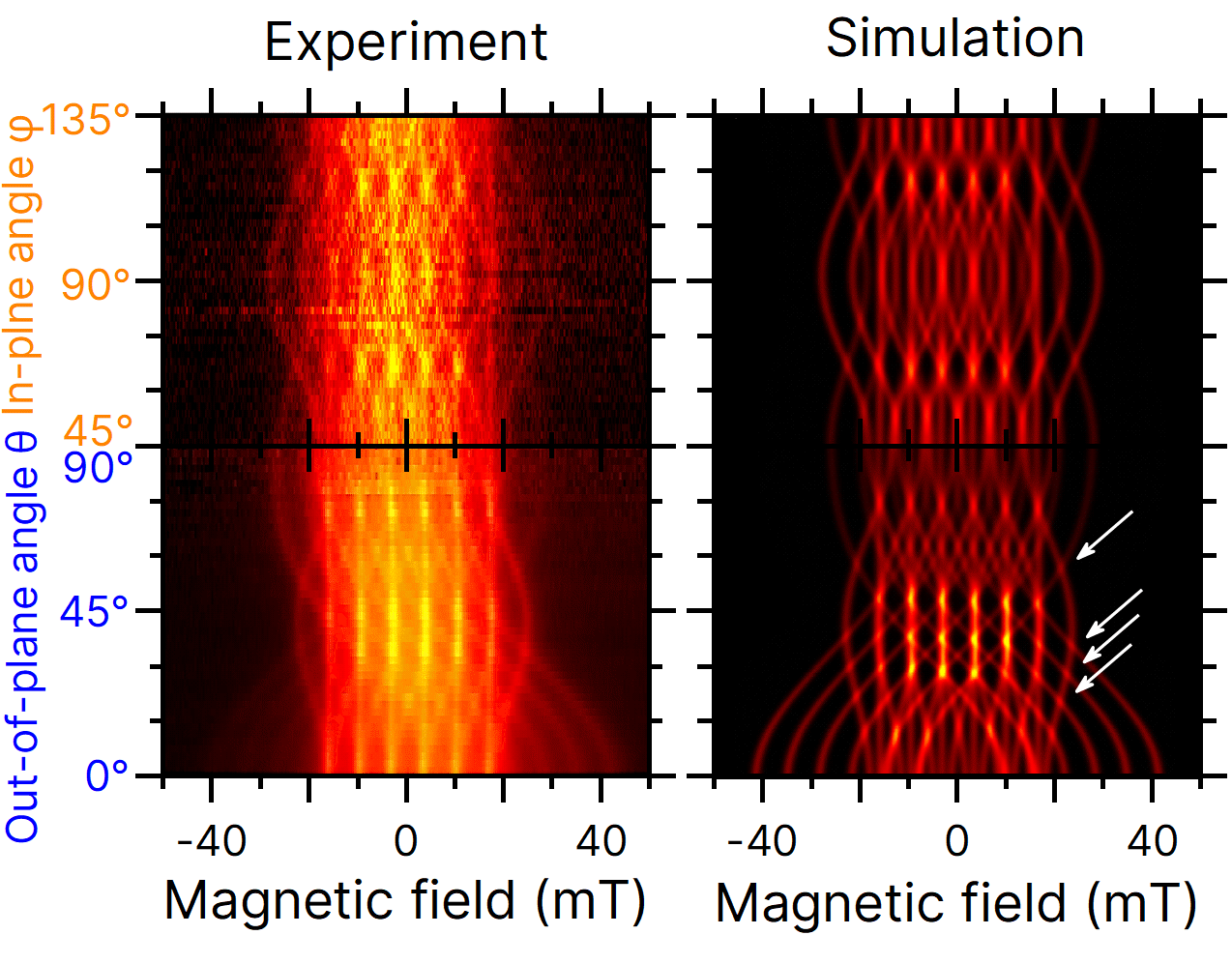}
\end{center}
\caption[]{(Color online) The left panel shows measured angular maps of the ODMR signal for the exemplary high-strain sample (UW0677). The intensity of the ODMR
signal is coded in the brightness of the map. Upper left part (in-plane
scan) was obtained for fixed out-of-plane angle $\theta=90^{\circ}$.
A lower left part (out-of-plane) scan was obtained for fixed in-plane angle
$\varphi=45^{\circ}$. The right part presents results of simulations obtained by numerical solving of the spin Hamiltonian presented in equation \ref{eq:hamiltonian}. Fitted spin Hamiltonian parameters are
$D=(497\pm 3)$~neV and $E=(-8\pm 13)$~neV.}
\label{fig:meduza_high_strain}
\vspace*{\floatsep}
\begin{center}
\includegraphics[width=85.9mm]{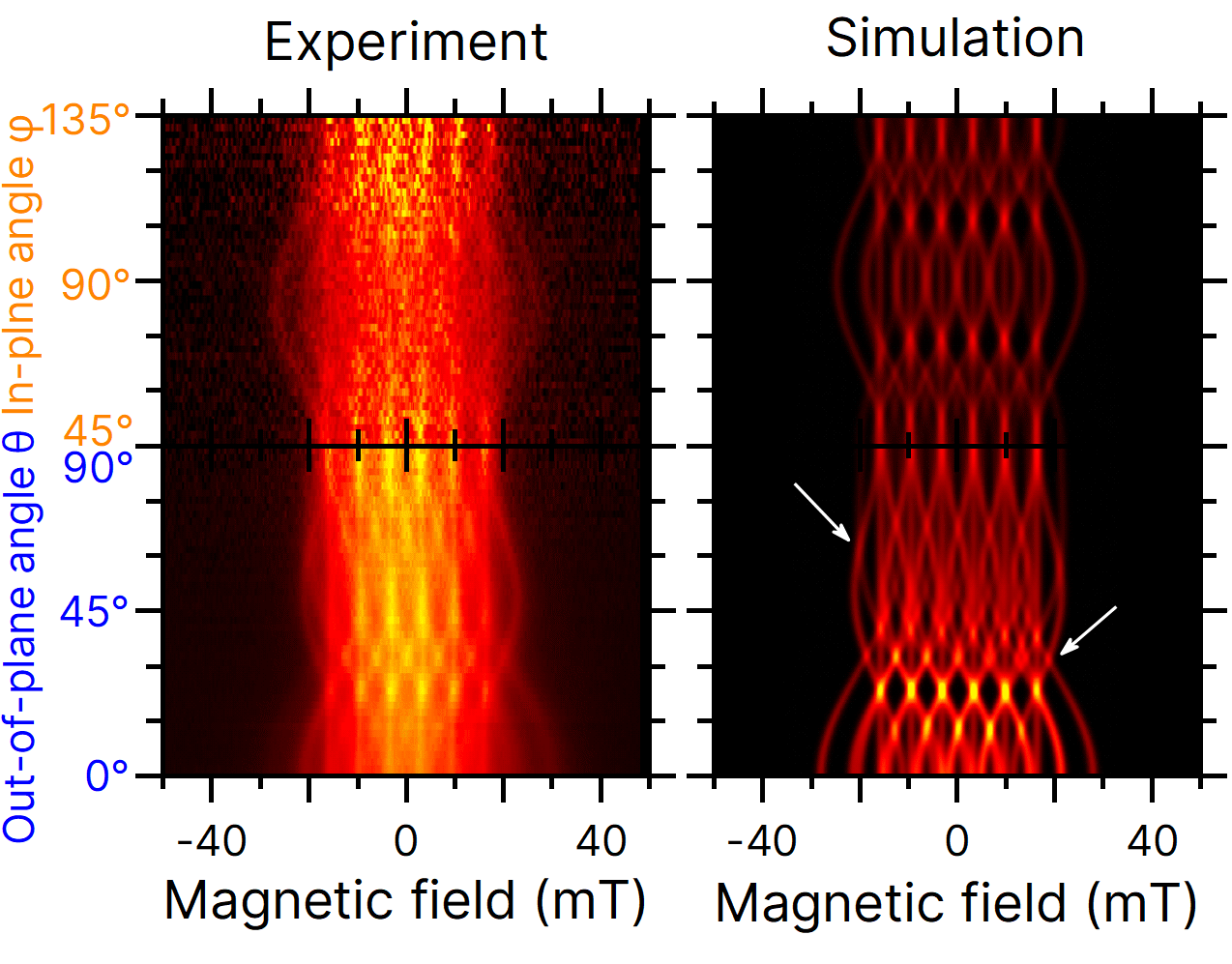}
\end{center}
\caption[]{(Color online) The left panel shows measured angular maps of the ODMR
signal for the exemplary low-strain sample (UW0676). The intensity of the ODMR signal is coded in the brightness of the map. The upper left part (in-plane scan) was obtained for fixed out-of-plane angle $\theta=90^{\circ}$. A lower left part (out-of-plane) scan was obtained for fixed in-plane angle
$\varphi=45^{\circ}$. The right part presents results of simulations obtained by numerical solving of the spin Hamiltonian presented in equation \ref{eq:hamiltonian}. Fitted spin Hamiltonian parameters are
$D=(155\pm 3)$~neV and $E=(0\pm 25)$~neV.}
\label{fig:meduza_low_strain}
\end{figure}

Angular resolved ODMR measurements can give precise information about spin
Hamiltonian parameters $D$ and $E$. While single ODMR measurement in Faraday
configuration should in principle give information about the $D$ parameter, the
angular ODMR can provide much more detailed information and increase experimental accuracy. The left panels of figure \ref{fig:meduza_high_strain} and
\ref{fig:meduza_low_strain} show angular ODMR measurements of example
high-strain and low-strain samples respectively. 

In the lower part of
the plot, the angular scan from the Faraday configuration
(magnetic field direction $\theta=0^{\circ}$) to the Voigt configuration
(magnetic field direction $\theta=90^{\circ}$) is presented. The upper part of the plot shows an in-plane ODMR angular scan starting from
$\langle$110$\rangle$ direction (parallel to the samples' edge) for
$\varphi=45^{\circ}$. The whole ODMR angular map resents a set of
characteristic features in which location on the map allows for the
determination of spin Hamiltonian parameters by comparison with similar
numerically simulated maps (see right panels).      

The first characteristic feature is the width of the ODMR spectrum in
Faraday configuration. In this configuration, a transition that is visible at the lowest magnetic field corresponds to the change of the electron spin projection from $S_z=+3/2$ to $S_z=+5/2$ with nuclear spin projection $I_z=+5/2$. Similarly, the transition at the highest magnetic field corresponds to the change of the electron spin projection from $S_z=-5/2$ to $S_z=-3/2$ with nuclear spin projection $I_z=-5/2$.  The measure of splitting between the outermost lines in units of the magnetic field directly corresponds to the value of the $D$ parameter in the spin Hamiltonian. At high magnetic field approximation -- assuming that eigenstates are the same as the eigenstates of the $z$ components of the spin operators, the spin Hamiltonian can be analytically solved. In such a case the $D$ parameter can be calculated as
\begin{equation}
D = (g_{\mbox{\tiny{Mn}}}\mu_B \Delta B_z -5A-4a)/8,
\label{eq:analytical_D}
\end{equation}
where $\Delta B_z$ is the distance in magnetic field units between the lowest and the highest transition lines in the ODMR signal. 

The above expression enables the determination of the $D$ parameter for $D\gtrapprox 75$~neV. Below $75$~neV, the transitions between other states are overlapping, and resolving the lines becomes cumbersome. The difference between $D$ value calculated from equation \ref{eq:analytical_D} for $75$~neV~$<D<1000$~neV and obtained from numerical solving of the spin Hamiltonian for $f=15.6$~GHz is below $0.05\%$. For higher values of $D$ the difference between the numerical solution and analytical approximation slowly increases but is still below $0.1\%$ for  $D$ as high as $3000$~neV.     

Moreover, the positions of the ODMR signal line crossings in ODMR angular maps (like those present between B$\approx+20$~mT to B$\approx+30$~mT at $\theta\approx30^{\circ}$ and $\theta\approx50^{\circ}$ in figure
\ref{fig:meduza_high_strain} or in figure \ref{fig:meduza_low_strain}, marked with white arrows)
enable for reducing fitting uncertainties down to a few neV. Similarly, by
analyzing analogous features of an in-plane angular scan we can fit the spin
Hamiltonian $E$ parameter which in the case of QWs should be zero. Indeed within fitting accuracy, the $E$ parameter for our samples is negligible.

\begin{figure}[h]
\begin{center}
\includegraphics[width=85.9mm]{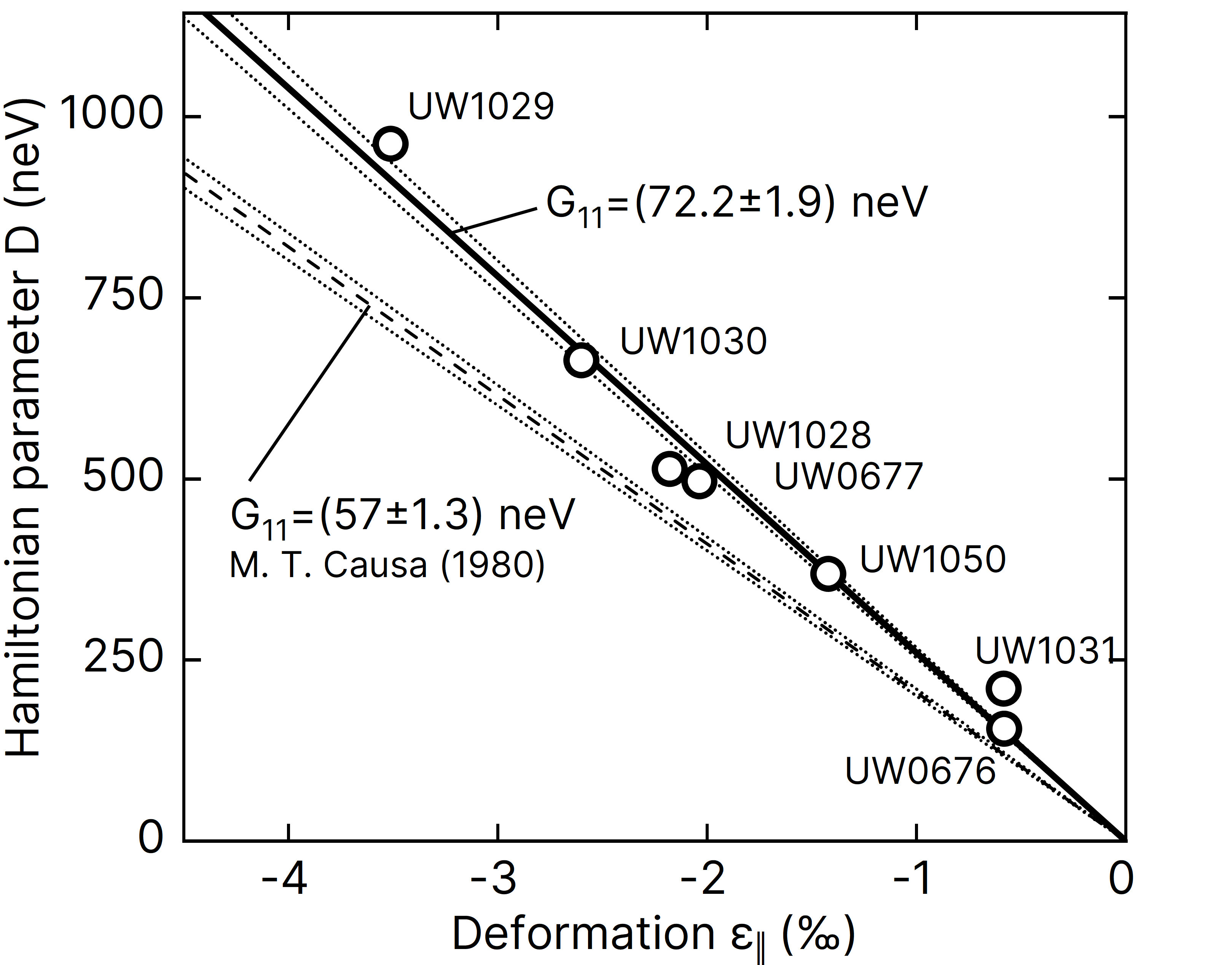}
\end{center}
\caption[]{The spin Hamiltonian $D$ parameter obtained from angular ODMR measurements as a function of deformation for analyzed quantum well samples (circles). The dashed curve represents the value of G$_{11}$ for bulk CdMnTe reported in Ref. \onlinecite{Causa_1980_PL}. The solid black curve represents the linear fit to data points measured in this work for Mn$^{2+}$ in CdTe and corresponds to \Gvalue. The dotted lines represent one standard error deviation. }
\label{fig:D_vs_epsilon}
\end{figure}

Figure \ref{fig:D_vs_epsilon} presents the obtained $D$ parameter as a function of deformation. As it is known \cite{Qazzaz_1995_SSC} the spin Hamiltonian $D$ parameter is related to deformation $\veps_{\parallel}$ by formula 
\begin{equation}
D=-\frac{3}{2} G_{11} \left( 1+ \frac{2 C_{12}}{C_{11}}\right) \veps_{\parallel}
\label{eq:D_G11_eps}
\end{equation}
 where $G_{11}$ is strain spin-lattice
coefficient and $C_{11}$, $C_{12}$ are elastic stiffness constants. The value of
$G_{11}$ strain spin-lattice coefficient obtained from our data is \Gvalue. This is significantly larger value than the only previously reported value\cite{Causa_1980_PL} of $G^{\mbox{\footnotesize Causa}}_{11}=(57\pm1.3)$~neV.


\section{Summary and conclusions}
We have presented a precise method of determination of spin Hamiltonian
parameters by angular resolved ODMR measurements. Due to the high sensitivity of ODMR measurements (we were probing
$\approx$3.8$\times$10$^9$ manganese spins which are almost 3 orders of
magnitude less than the limit for standard EPR setup) we obtained strain-induced axial-symmetry spin Hamiltonian parameter $D$ for single-layered quantum wells of 10~nm thickness with neV accuracy. By determination of $D$ values for a series of samples differing by
magnesium content we were able to determine strain spin-lattice coefficient \Gvalue ~for Mn$^{2+}$ ion in CdTe. 
\section{Acknowledgements}

This work was supported by the Polish National Science Centre under decisions DEC-2016/23/B/ST3/03437
and DEC-2015/18/E/ST3/00559. It has also received funding from the Norwegian Financial Mechanism 2014-2021
within project No. 2020/37/K/ST3/03656 and from the Polish National Agency for Academic Exchange within Polish Returns program under Grant No. PPN/PPO/2020/1/00030.
The project was carried out with the use of CePT, CeZaMat, and NLTK infrastructures financed by the European Union - the European Regional Development Fund within the Operational Programme “Innovative economy” for 2007 - 2013. One of us (P.K.) has been supported by the ATOMOPTO project (TEAM programme of the Foundation for Polish Science, cofinanced by the EU within the ERDFund).

\appendix
\section{Temperature dependence of the thermal expansion coefficients of CdTe and GaAs}
\label{app:thermal_coeff}
The substrate material of analyzed samples is made of GaAs on top of which a thick CdTe layer is deposited. As the CdTe is deposited at a temperature of around 560~K, the XRD measurements are performed at 300~K and optical measurements at 1.6~K it is crucial to take into account the mechanical properties of the samples. In fact, the presented design is an example of composite-material behaving similarly to the bimetallic strip, as the thermal expansion coefficients $\alpha$ of GaAs and CdTe are different. Moreover, the changes of thermal expansion coefficient of both materials -- $\alpha$(T), as the temperature changes, are different. As a result, accumulated deformation, due to the cooling, must be evaluated based on full $\alpha$(T) curves. The values of $\alpha$ are available in the literature for both CdTe\cite{Novikova_1960_FTT,Williams_1969_SSC,Browder_1972_AOA,Novikova_1974_book,Bagot_1993_PSSB} and GaAs\cite{Sparks_1967_PR,Feder_1968_JoAP,Feder_1968_JoAP,Novikova_1974_book,Soma_1982_SSC,Sirota_1984_DANS}. The additional deformation present in CdTe layer can be then calculated as
\begin{equation}
\Delta\varepsilon_{\parallel}^{\textrm{temp.}}=\int_{T_0}^{T_i}\left(\alpha_{\textrm{GaAs}}(T)-\alpha_{\textrm{CdTe}}(T)\right)dT
\label{eq:delta_epsilon_temperaturowy}
\end{equation}
For convinient integration we interpolated literature data using an empirical formula:
\begin{equation}
\alpha(T) \approx \sum_{i=1}^{i=3} G_i(g_i,s_i,T)
\label{eq:thermal}
\end{equation}
where 
\begin{equation}
G_i(g_i,s_i,T)=g_i\left(\frac{s_i}{T}\right)^2\frac{e^{\frac{s_i}{T}}}{(e^{\frac{s_i}{T}}-1)^2}
\label{eq:thermal_part}
\end{equation}
Full theory and formulas are presented elsewhere\cite{Bagot_1993_PSSB}. However, only the combination of $\alpha$ curves from ref. \onlinecite{Bagot_1993_PSSB} for CdTe with ref. \onlinecite{Novikova_1974_book} GaAs gives deformation value that agrees with values obtained from XRD measurements. As the temperature of the substrate during the MBE growth of the sample was known within 10~K accuracy we evaluate actual growth temperature $T_0$ as a value that results in the $\varepsilon_{\parallel}=-0.389\prom$ measured with XRD. This procedure gives $T_0=552$~K.

\begin{figure}[h]
\begin{center}
\includegraphics[width=85.9mm]{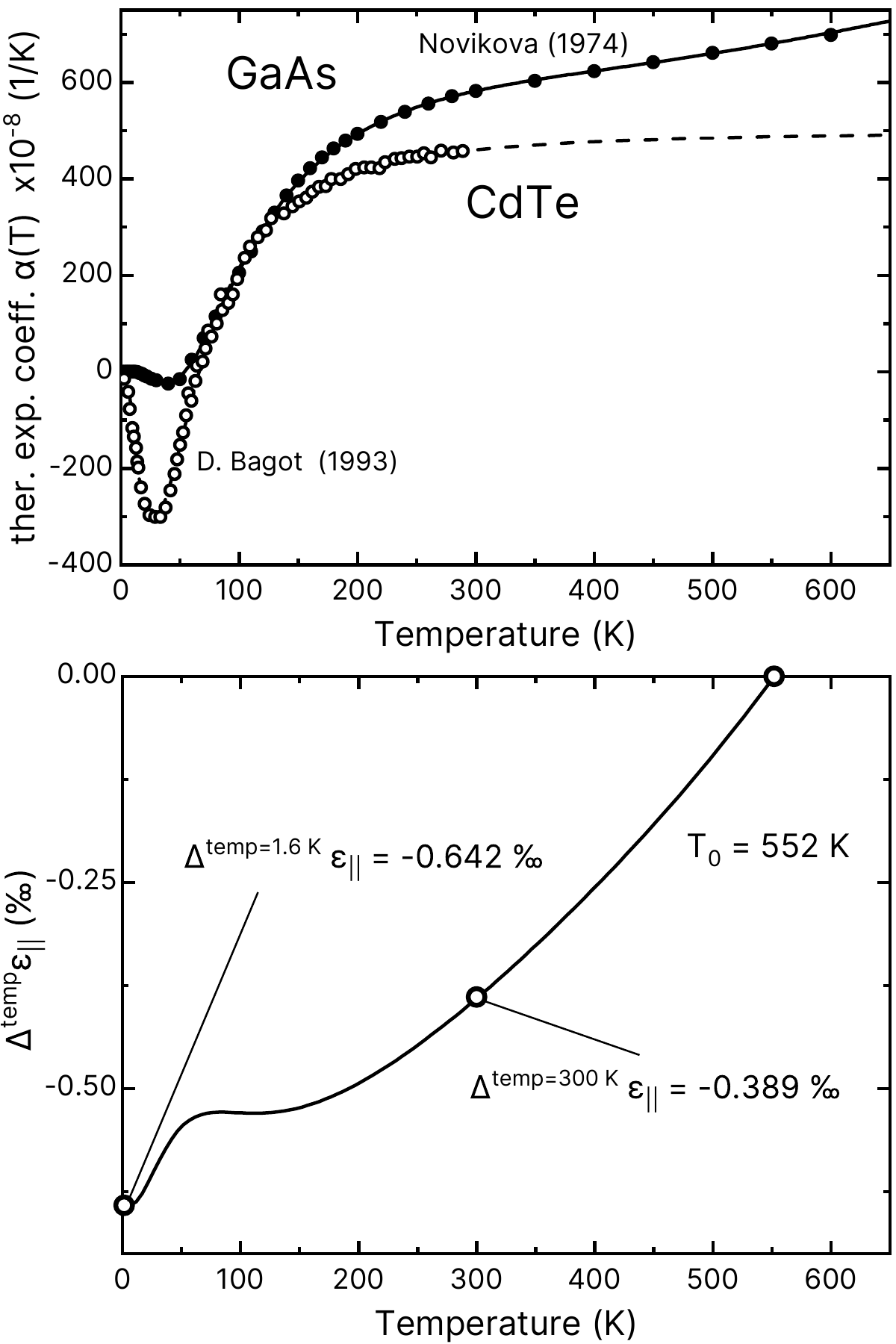}
\end{center}
\caption[]{The upper panel presents linear thermal expansion coefficients of CdTe and
GaAs as a function of temperature. Data for CdTe and GaAs were taken from
papers of Bagot et al.\cite{Bagot_1993_PSSB} and Novikova et
al.\cite{Novikova_1974_book} respectively. Curves are best-fit results of
fitting equation \ref{eq:thermal} to the data. The lower panel shows calculation results of the additional deformation $\Delta\varepsilon_{\parallel}^{\textrm{temp.}}$ present in the QW samples due to the difference in thermal expansion coefficients of CdTe material and GaAs substrate material. The $T_0$ is the temperature related to the MBE growth process.}
\label{fig:app_thermal}
\end{figure}

\begin{table}[]
\caption{Numerical values of s$_{i}$, g$_{i}$ coefficients used for reproduction of the linear thermal expansion coefficients $\alpha(T)$ of CdTe and GaAs with eq. \ref{eq:thermal}.}
\begin{tabular}{lcccS[table-format=7.5]S[table-format=5.5]}
\toprule
material  &        & i & & {s$_{i}$~(K)}   & {g$_{i}$~(K$^{-1}$)}   \\ \hline\hline
 CdTe    &        & 1 & & 222.02098    & 883.96417     \\
         &        & 2 & & -40.90023    & -384.86715    \\
         &        & 3 & & -13241.61    & 1403.50696    \\
\cline{2-6} 
 GaAs    &        & 1 & & 3562.24838   & 640.97475     \\
         &        & 2 & & -99.54697    & -44.50516     \\
         &        & 3 & & 371.47981    & 709.90672     \\ \hline \hline
\end{tabular}
\end{table}

\section{Temperature dependence of the CdTe elastic stiffness constants}
R. D. Greenough and S. B. Palmer in ref. \onlinecite{Greenough_1973_JPAP} presented experimentally determined linear combinations of the CdTe elastic parameters values from 300~K to 4.2~K. From their data one can calculate $2C_{12}/C_{11}$ ratio.  In this paper we evaluate $2C_{12}/C_{11}$ ratio using 2 components of equation \ref{eq:thermal} and added constant $C_0$. We obtain  the best reproduction of data from \onlinecite{Greenough_1973_JPAP} for ~g$_{1}=-0.0092$~K$^{-1}$, ~g$_{2}=-0.015$~K$^{-1}$, ~s$_{1}=-516$~K, ~s$_{2}=1415$~K and~$C_0=1.39886$. For temperatures below 50~K, as the temperature decreases, the value of $C_0$ approaches the $2C_{12}/C_{11}$ ratio. For 300~K and 1.6~K we find  $2C_{12}/C_{11}|_{300~\mathrm{K}}=1.38859$ and $2C_{12}/C_{11}|_{1.6~\mathrm{K}}=1.39886$.

\section{Numerical simulation of QW levels}
Theoretical calculations of energy levels in QWs were performed by solving a 1D Schrodinger equation of carrier confined in the square quantum well. The effective masses in perpendicular direction are assumed $0.1m_0$, $0.63m_0$, and $0.13m_0$ for electron, heavy and light hole, respectively where $m_0$ is the free electron mass. The potential was obtained from composition profiles. The QW layer was composed of $\mathrm{Cd}_{1-y}\mathrm{Mn}_y\mathrm{Te}$, sandwiched between the $\mathrm{Cd}_{1-x}\mathrm{Mg}{_x}\mathrm{Te}$ barriers, the chemical potential $A_C$ and $A_V$ for conduction and valence band is defined by the energy gap difference with a relative valence band offset $Q_V = 0.4$ \cite{Wojtowicz_1996_APL}.

We used bandgap versus composition relation\cite{Gaj_2010_book}:

\begin{equation}
    E_g^{\mathrm{Cd}_{1-x}\mathrm{Mg}_x\mathrm{Te}}(x) = B x + E_g^{\mathrm{CdTe}} = (1.850 x + 1.606) \mathrm{~eV} 
    \label{equation:Eg3}
\end{equation}

for the barrier material and for the QW:
\begin{equation}
    E_g^{\mathrm{Cd}_{1-y}\mathrm{Mn}_y\mathrm{Te}}(y) = B y + E_g^{\mathrm{CdTe}} = (1.563 y + 1.606) \mathrm{~eV} .
     \label{equation:Eg4}
\end{equation}

The difference of lattice constants of buffer and barrier layers \cite{Dynowska_1999_JAC} leads to the presence of additional deformation potential $A_{i}^{\mathrm{deform}} = \alpha_i \varepsilon_{\parallel}$, where $i = \mathrm{E,HH,LH}$ denotes electron, heavy and light hole. The $\alpha_i$ parameters are corresponding to the $a'$ (hydrostatic deformation potential) and $b$ (shear deformation potential) material constants and elastic constants ratio. For zinc blend crystals, as CdTe, the modified potential depths for the electrons, heavy and light holes are \cite{Chuang_2009_book}:

\begin{equation}
    A_{\mathrm{E}} = A_C + A_{\mathrm{E}}^{\mathrm{deform}} = A_C + \frac{2a'}{3} \left (2 - \frac{2 C_{12}}{C_{11}} \right ) \varepsilon_{\parallel} ,
    \label{equation:Eg1}
\end{equation}

\begin{eqnarray}
\nonumber A_{\mathrm{HH,LH}}= A_V + A_{\mathrm{HH,LH}}^{\mathrm{deform}} = \, \, \, \, \\ 
\label{label} = A_V +  \left[-\frac{a'}{3} \left (2 - \frac{2 C_{12}}{C_{11}} \right )  \pm b \left (1+\frac{2C_{12}}{C_{11}} \right) \right ] \varepsilon_{\parallel}.\quad
\end{eqnarray}

The elastic constants $C_{12}$ and $C_{11}$ are taken from literature \cite{Greenough_1973_JPAP}, for this calculation we have used  \Cratio, while the parameter $a' = -3.85$~eV \cite{Peyla_1992_PRB}. The light - heavy hole potential difference corresponds only to the difference in the deformation potential (see equation \ref{eq:Delta_LHHH}).

The shear deformation potential \bvalue ~was obtained from the linear fit to the measured light-heavy hole splitting versus calculated $\varepsilon_{\parallel}$ as it was mentioned before. The exciton binding energy for heavy hole exciton was set as $18$~meV, while for light hole exciton $21$~meV, evaluated from the literature \cite{Mathieu_1992_PRB,Christol_1993_JoAP,Gaj_1994_PRB}. The QW width was equal to $10.45$~nm, the same for each studied sample. With these assumptions, we have calculated heavy and light hole exciton energy levels and their energy difference presented in Fig. \ref{fig:HH_LH_eps}. Finally, the deformation $\varepsilon_{\parallel}$ splits heavy and light hole exciton energy levels. 

Here we assume that the other parameters as QW width and exciton binding energy are not vulnerable to the deformation. Moreover, the QW width, deformation potential corresponding to the $a'$ parameter, and the exciton binding energy affect the exciton energy in a similar way - increasing or decreasing the state energy. Nonetheless, this additional energy offset does not affect the heavy-light hole splitting strain dependence which we are interested in here.

\bibliography{Bibliografia_ODMR}

\end{document}